\documentclass[12pt,a4paper]{article}

\usepackage{ifthen} 
\newboolean{pdflatex}
\setboolean{pdflatex}{true} 

\newboolean{articletitles}
\setboolean{articletitles}{true} 

\newboolean{uprightparticles}
\setboolean{uprightparticles}{false} 

\newboolean{inbibliography}
\setboolean{inbibliography}{false} 

\def\paperauthors{D. Mueller, M. Clemencic, G. Corti, M. Gersabeck} 
\def\paperasciititle{ReDecay: A novel approach to speed up the simulation at LHCb} 
\def\papertitle{ReDecay: A novel approach to speed up the simulation at \lhcb}
\def\paperkeywords{{High Energy Physics}, {Fast simulation}, {LHCb}} 
\def\papercopyright{\the\year\ CERN for the benefit of the LHCb collaboration} 
\def\paperlicence{CC-BY-4.0 licence}
\def\paperlicenceurl{https://creativecommons.org/licenses/by/4.0/}

\usepackage[top=1in, bottom=1.25in, left=1in, right=1in]{geometry}

\usepackage{microtype}
\usepackage{lineno}  
\usepackage{xspace} 
\usepackage{caption} 

\usepackage{graphicx}  
\usepackage{color}
\usepackage{colortbl}
\graphicspath{{./figs/}} 

\usepackage{amsmath} 
\usepackage{amssymb}
\usepackage{amsfonts}
\usepackage{upgreek} 

\newcommand*\patchAmsMathEnvironmentForLineno[1]{%
\expandafter\let\csname old#1\expandafter\endcsname\csname #1\endcsname
\expandafter\let\csname oldend#1\expandafter\endcsname\csname
end#1\endcsname
 \renewenvironment{#1}%
   {\linenomath\csname old#1\endcsname}%
   {\csname oldend#1\endcsname\endlinenomath}%
}
\newcommand*\patchBothAmsMathEnvironmentsForLineno[1]{%
  \patchAmsMathEnvironmentForLineno{#1}%
  \patchAmsMathEnvironmentForLineno{#1*}%
}
\AtBeginDocument{%
\patchBothAmsMathEnvironmentsForLineno{equation}%
\patchBothAmsMathEnvironmentsForLineno{align}%
\patchBothAmsMathEnvironmentsForLineno{flalign}%
\patchBothAmsMathEnvironmentsForLineno{alignat}%
\patchBothAmsMathEnvironmentsForLineno{gather}%
\patchBothAmsMathEnvironmentsForLineno{multline}%
\patchBothAmsMathEnvironmentsForLineno{eqnarray}%
}

\usepackage{hyperxmp}

\usepackage[pdftex,
            pdfauthor={\paperauthors},
            pdftitle={\paperasciititle},
            pdfkeywords={\paperkeywords},
            pdfcopyright={Copyright (C) \papercopyright},
            pdflicenseurl={\paperlicenceurl}]{hyperref}

\usepackage[all]{hypcap} 

\usepackage{xspace} 
\usepackage{upgreek}

\def\lhcb {\mbox{LHCb}\xspace}

\def\MagUp {\mbox{\em Mag\kern -0.05em Up}\xspace}

\ifthenelse{\boolean{uprightparticles}}%
{

 \def\Pnu         {\ensuremath{\upnu}\xspace}                 
                  
 \def\Ppi         {\ensuremath{\uppi}\xspace}

 \def\Ptau        {\ensuremath{\uptau}\xspace}

 \def\PDelta      {\ensuremath{\Delta}\xspace}                 
 \def\PXi      {\ensuremath{\Xi}\xspace}                 
 \def\PLambda      {\ensuremath{\Lambda}\xspace}                 
 \def\PSigma      {\ensuremath{\Sigma}\xspace}                 
 \def\POmega      {\ensuremath{\Omega}\xspace}                 
 \def\PUpsilon      {\ensuremath{\Upsilon}\xspace}                 
 
 \def\PB      {\ensuremath{\mathrm{B}}\xspace}                 
                  
 \def\PD      {\ensuremath{\mathrm{D}}\xspace}

 \def\PK      {\ensuremath{\mathrm{K}}\xspace}

 \def\Pi      {\ensuremath{\mathrm{i}}\xspace}

}
{

 \def\Pnu         {\ensuremath{\nu}\xspace}                 
                  
 \def\Ppi         {\ensuremath{\pi}\xspace}

 \def\Ptau        {\ensuremath{\tau}\xspace}

 \mathchardef\PDelta="7101
 \mathchardef\PXi="7104
 \mathchardef\PLambda="7103
 \mathchardef\PSigma="7106
 \mathchardef\POmega="710A
 \mathchardef\PUpsilon="7107
                  
 \def\PB      {\ensuremath{B}\xspace}                 
                  
 \def\PD      {\ensuremath{D}\xspace}

 \def\PK      {\ensuremath{K}\xspace}

 \def\Pi      {\ensuremath{i}\xspace}

}

\makeatletter
\ifcase \@ptsize \relax
  \newcommand{\miniscule}{\@setfontsize\miniscule{4}{5}}
\or
  \newcommand{\miniscule}{\@setfontsize\miniscule{5}{6}}
\or
  \newcommand{\miniscule}{\@setfontsize\miniscule{5}{6}}
\fi
\makeatother

\DeclareRobustCommand{\optbar}[1]{\shortstack{{\miniscule (\rule[.5ex]{1.25em}{.18mm})}
  \\ [-.7ex] $#1$}}

\def\taup       {{\ensuremath{\Ptau^+}}\xspace}

\def\neu        {{\ensuremath{\Pnu}}\xspace}

\def\neut       {{\ensuremath{\neu_\tau}}\xspace}

\def\pion   {{\ensuremath{\Ppi}}\xspace}

\def\pip    {{\ensuremath{\pion^+}}\xspace}
\def\pim    {{\ensuremath{\pion^-}}\xspace}

\def\kaon    {{\ensuremath{\PK}}\xspace}
  \def\Kbar    {{\kern 0.2em\overline{\kern -0.2em \PK}{}}\xspace}

\def\KorKbar    {\kern 0.18em\optbar{\kern -0.18em K}{}\xspace}

\def\Kp      {{\ensuremath{\kaon^+}}\xspace}
\def\Km      {{\ensuremath{\kaon^-}}\xspace}

  \def\Dbar    {{\kern 0.2em\overline{\kern -0.2em \PD}{}}\xspace}
\def\D       {{\ensuremath{\PD}}\xspace}

\def\DorDbar    {\kern 0.18em\optbar{\kern -0.18em D}{}\xspace}
\def\Dz      {{\ensuremath{\D^0}}\xspace}

\def\Dstarm  {{\ensuremath{\D^{*-}}}\xspace}

\def\B       {{\ensuremath{\PB}}\xspace}
\def\Bbar    {{\ensuremath{\kern 0.18em\overline{\kern -0.18em \PB}{}}}\xspace}

\def\BorBbar    {\kern 0.18em\optbar{\kern -0.18em B}{}\xspace}
\def\Bz      {{\ensuremath{\B^0}}\xspace}

  \def\Y#1S{\ensuremath{\PUpsilon{(#1S)}}\xspace}

\def\Lbar        {{\ensuremath{\kern 0.1em\overline{\kern -0.1em\PLambda}}}\xspace}
\def\LorLbar    {\kern 0.18em\optbar{\kern -0.18em \PLambda}{}\xspace}

\def\AT#1     {\ensuremath{A_{\mathrm{T}}^{#1}}\xspace}           

\def\C#1      {\ensuremath{\mathcal{C}_{#1}}\xspace}                       
\def\Cp#1     {\ensuremath{\mathcal{C}_{#1}^{'}}\xspace}                    
\def\Ceff#1   {\ensuremath{\mathcal{C}_{#1}^{\mathrm{(eff)}}}\xspace}        
\def\Cpeff#1  {\ensuremath{\mathcal{C}_{#1}^{'\mathrm{(eff)}}}\xspace}       
\def\Ope#1    {\ensuremath{\mathcal{O}_{#1}}\xspace}                       
\def\Opep#1   {\ensuremath{\mathcal{O}_{#1}^{'}}\xspace}                    

\newcommand{\tev}{\ifthenelse{\boolean{inbibliography}}{\ensuremath{~T\kern -0.05em eV}}{\ensuremath{\mathrm{\,Te\kern -0.1em V}}}\xspace}
\newcommand{\gev}{\ensuremath{\mathrm{\,Ge\kern -0.1em V}}\xspace}
\newcommand{\mev}{\ensuremath{\mathrm{\,Me\kern -0.1em V}}\xspace}
\newcommand{\kev}{\ensuremath{\mathrm{\,ke\kern -0.1em V}}\xspace}
\newcommand{\ev}{\ensuremath{\mathrm{\,e\kern -0.1em V}}\xspace}
\newcommand{\gevc}{\ensuremath{{\mathrm{\,Ge\kern -0.1em V\!/}c}}\xspace}
\newcommand{\mevc}{\ensuremath{{\mathrm{\,Me\kern -0.1em V\!/}c}}\xspace}
\newcommand{\gevcc}{\ensuremath{{\mathrm{\,Ge\kern -0.1em V\!/}c^2}}\xspace}
\newcommand{\gevgevcccc}{\ensuremath{{\mathrm{\,Ge\kern -0.1em V^2\!/}c^4}}\xspace}
\newcommand{\mevcc}{\ensuremath{{\mathrm{\,Me\kern -0.1em V\!/}c^2}}\xspace}

\def\gsim{{~\raise.15em\hbox{$>$}\kern-.85em
          \lower.35em\hbox{$\sim$}~}\xspace}
\def\lsim{{~\raise.15em\hbox{$<$}\kern-.85em
          \lower.35em\hbox{$\sim$}~}\xspace}

\def\evtgen     {\mbox{\textsc{EvtGen}}\xspace}

\def\gaudi      {\mbox{\textsc{Gaudi}}\xspace}
\def\gauss      {\mbox{\textsc{Gauss}}\xspace}
\def\geant      {\mbox{\textsc{Geant4}}\xspace}

\def\photos     {\mbox{\textsc{Photos}}\xspace}

\def\pythia     {\mbox{\textsc{Pythia}}\xspace}

\def\tell1  {TELL1\xspace}
\def\ukl1   {UKL1\xspace}

\newcommand{\eg}{\mbox{\itshape e.g.}\xspace}

\usepackage{cite} 
\usepackage{mciteplus}

\usepackage{longtable} 
\usepackage{subcaption}
\usepackage[textsize=tiny]{todonotes}
\usepackage[capitalise, noabbrev]{cleveref}

\begin{document}

\renewcommand{\thefootnote}{\fnsymbol{footnote}}
\setcounter{footnote}{1}


\begin{titlepage}
\pagenumbering{roman}

\vspace*{-1.5cm}
\centerline{\large EUROPEAN ORGANIZATION FOR NUCLEAR RESEARCH (CERN)}
\vspace*{1.5cm}
\noindent
\begin{tabular*}{\linewidth}{lc@{\extracolsep{\fill}}r@{\extracolsep{0pt}}}
\ifthenelse{\boolean{pdflatex}}
{\vspace*{-1.5cm}\mbox{\!\!\!\includegraphics[width=.14\textwidth]{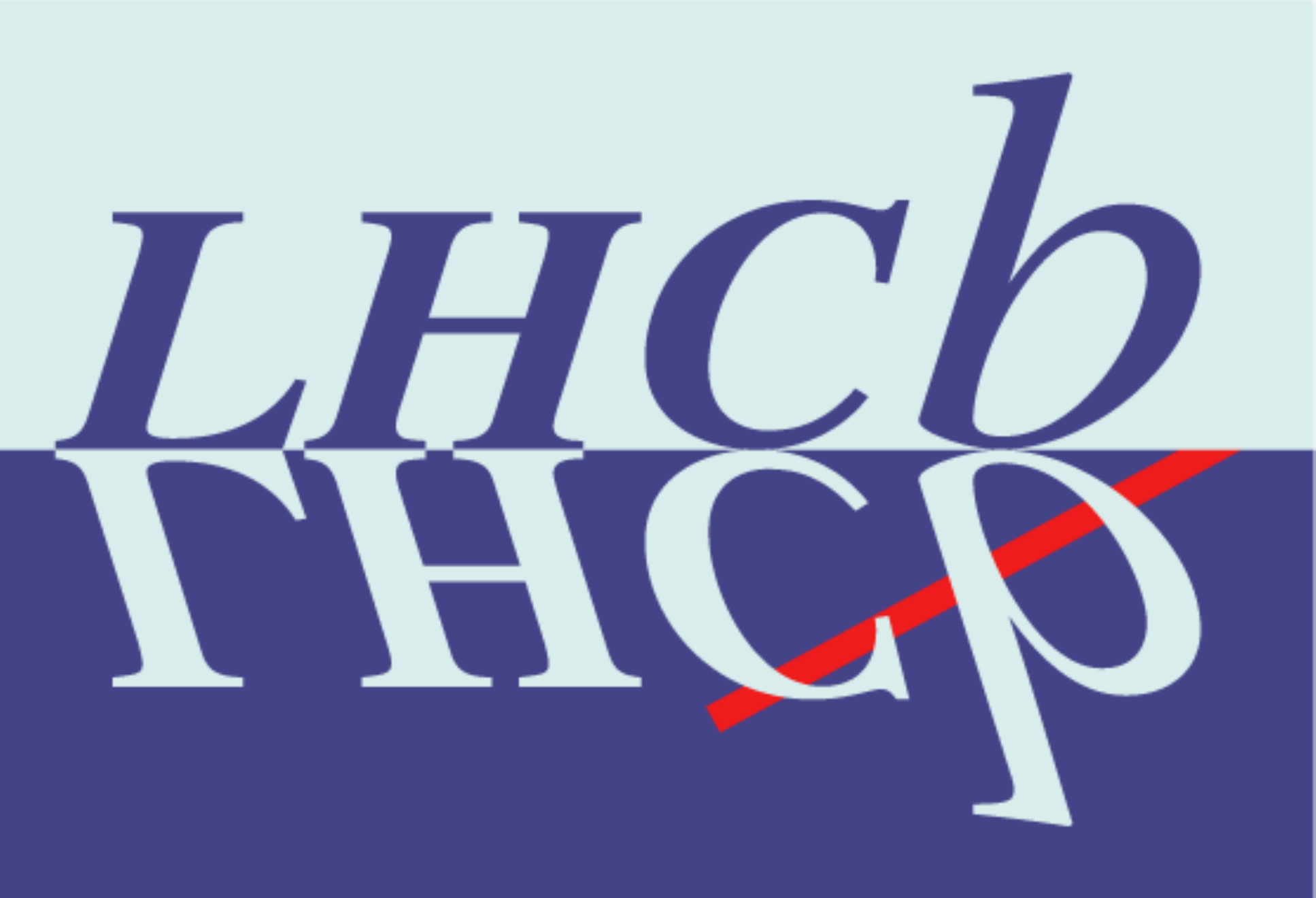}} & &}%
{\vspace*{-1.2cm}\mbox{\!\!\!\includegraphics[width=.12\textwidth]{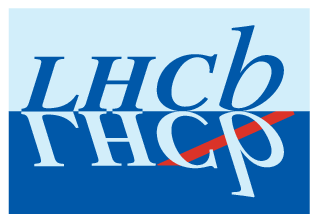}} & &}%
\\
 & & CERN-LHCb-DP-2018-004 \\  
 & & December 14, 2018 \\ 
\end{tabular*}

\vspace*{3.0cm}

{\normalfont\bfseries\boldmath\huge
\begin{center}
  \papertitle 
\end{center}
}

\vspace*{1.0cm}

\begin{center}
  D. M\"uller$^{1}$, M. Clemencic$^{1}$, G. Corti$^{1}$, M. Gersabeck$^{2}$

\bigskip{\it
\footnotesize
$ ^{1}$European Organization for Nuclear Research (CERN), Geneva, Switzerland\\
$ ^{2}$School of Physics and Astronomy, University of Manchester, Manchester, United Kingdom\\
}
\end{center}

\vspace{\fill}

\begin{abstract}
\noindent With the steady increase in the precision of flavour physics measurements collected during LHC Run 2, 
the LHCb experiment requires simulated data samples of larger and larger sizes to study the detector response in detail. 
The simulation of the detector response is the main contribution to the time needed to simulate full events.
This time scales linearly with the particle multiplicity. 
Of the dozens of particles present in the simulation only the few participating in the signal decay under study are of interest, 
while all remaining particles mainly affect the resolutions and efficiencies of the detector.
This paper presents a novel development for the LHCb simulation software which re-uses the rest of the event from previously simulated events. 
This approach achieves an order of magnitude increase in speed and the same quality compared to the nominal simulation.

\end{abstract}

\vspace*{2.0cm}

\begin{center}
  Published in Eur.~Phys.~J.~C \textbf{78} (2018) 1009
\end{center}

\vspace{\fill}

{\footnotesize 
\centerline{\copyright~\papercopyright. \href{\paperlicenceurl}{\paperlicence}.}}

\vspace*{2mm}

\end{titlepage}


\newpage
\setcounter{page}{2}
\mbox{~}
%
%
%
%


\renewcommand{\thefootnote}{\arabic{footnote}}
\setcounter{footnote}{0}



\pagestyle{plain} 
\setcounter{page}{1}
\pagenumbering{arabic}


\section{Introduction}
\label{sec:Introduction}

A common challenge in many measurements performed in high-energy physics is the necessity
to understand the effects of the detector response on the physics parameters of interest. This response
is driven by resolution effects that distort the true distribution of a quantity and by inefficiencies that are
introduced by either an imperfect reconstruction in the detector or a deliberate event selection, resulting in an
unknown number of true events for a given number of reconstructed events.
The extraction of an unbiased estimate of the physics parameter of interest requires the correction of these effects.
However, due to increasing complexity of the detectors employed and the challenging experimental conditions -- especially
at a hadron collider -- the only feasible solution is the generation of Monte Carlo (MC) events and the simulation of their
detector response to study the evolution from the generated to the reconstructed and selected objects.
In the era of the Large Hadron Collider, the simulation of the large event samples occupies a considerable fraction
of the overall computing resources available.
Hence, measures to decrease the time required to simulate an event are crucial to fully exploit the large
datasets recorded by the detectors. In fact, recent measurements hinting at tensions with respect to the
predictions of the Standard Model of elementary particle physics have systematic uncertainties that are dominated by the insufficient amount of
simulated data~\cite{LHCb-PAPER-2017-017}. 
Therefore, fast simulation options are necessary to further improve measurements such as these.

This paper presents a fast simulation approach that is applicable if the signal process is a decay of a heavy particle, \eg $\Dz\rightarrow\Km\pip$ or $\Bz\rightarrow\Dstarm\taup\neut$. It retains the same
quality as the nominal full simulation while reaching increases in speed of an order of magnitude. 
In \cref{sec:procedure}, the ReDecay approach is presented while \cref{sec:statistics} focuses on the correct treatment of correlations arising
in this approach. Lastly, \cref{sec:experience} discusses the experience with applying the approach within the \lhcb collaboration. While the approach
is applicable in all experiments which study the type of final state as defined above, the \lhcb experiment is used as an example throughout this paper for illustrative purposes.

\section{The ReDecay approach}
\label{sec:procedure}

In each simulated proton-proton collision, hundreds of particles are typically generated and tracked through a simulation
of the detector, most of the time based on the \geant toolkit~\cite{Allison:2006ve, *Agostinelli:2002hh}. Generally, the time needed to create an MC event for LHC experiments is dominated by the detector transport,
accounting for $95\%$ to $99\%$ of the total time, which is proportional to the number of particles that need to be tracked.
In the case that the intended measurement studies decays of heavy particles to exclusive final states with individual children being reconstructed for each particle,
all long lived particles in the event can be split into two distinct groups: the particles that participate in the
signal process and all remaining particles where the latter group is henceforth referred to as the rest of the event (ROE).
As the final state of the signal process is usually comprised of only a few particles, most
particles in the event are part of the ROE and hence the majority
of the computing time per event is spent on the simulation of particles that are never explicitly looked at.
Ideally, this situation should be inverted and most of the computing resources should be spent on simulating the signal decays themselves.
However, simply simulating the signal process without any contributions from the ROE results in a much lower occupancy of the detector.
This lower occupancy in turn leads to a significant mis-modelling of the detector response compared to the real detector: resolution effects are underestimated while reconstruction efficiencies are overestimated.

In the ReDecay approach, problems with simulating only the signal process are
mitigated by re-using the ROE multiple times instead of generating a new one
for every event. The signal particle is kept identical in all sets of events that use the same
ROE, to preserve correlations, but is independently
decayed each time. To be more precise, in each redecayed event, the origin vertex and
kinematics of the signal particle are identical, while the decay time and thus decay vertex as well
as the final state particle kinematics are different. 
The following procedure is applied:
\begin{enumerate}
  \item A full MC event including the signal decay is generated.
  \item Before the generated particles are passed through the detector simulation, the signal and its decay products are removed from the event
    and the origin-vertex position and momentum of the signal particle are stored.
  \item The remaining ROE is simulated as usual and the entire output, the information on the
    true particles and the energy deposits in the detector, are kept.
  \item A signal decay is generated and simulated using the stored origin vertex position and
    momentum.
  \item The persisted ROE and the signal decay are merged and written out to disk as a
    full event.
  \item The steps 4 and 5 are repeated $N_{\text{ReDecay}}$ times, where $N_{\text{ReDecay}}$ is a fixed number that is the same for all events.
\end{enumerate}
An ROE and a signal decay associated to form a complete event according to the procedure above are digitised simultaneously. This ensures that simulated energy deposits that stem from particles in the ROE can interfere with the deposits produced by the signal decay products, as is the case in the standard method to simulate events. Different complete events, for example obtained combining the same ROE with different signal decays, are themselves digitised independently.
This further implies that each event obtained could have been produced by chance in a nominal
simulation as ReDecay just reorders the already factorised approach: hadrons are decayed
independently and the quasi-stable tracks are propagated through the detector individually. Therefore, the
efficiencies and resolutions are, by construction, identical to those found in a nominal simulation.
Furthermore, with increasing $N_{\text{ReDecay}}$, the average time per created event becomes more and more dominated
by the time required to simulate the detector response for the signal particle and its decay products compared to the simulation of the particles from the ROE.
However, an attentive reader will have noticed that different events stemming from the same original event are correlated, \eg they all
have --- bar resolution effects --- the same signal kinematics. The magnitude of these correlations will also depend on the studied observables
and the following section attempts to quantify the strength of the correlations. Additionally, it presents a method to take them into account in 
an analysis using simulated samples that have been generated following the ReDecay approach.

\section{The statistical treatment}
\label{sec:statistics}
Re-using the ROE multiple times and leaving the kinematics of the signal particle unchanged yields correlated events.
Considering the signal decay $\Dz\rightarrow\Km\pip$ as an example large correlations are expected for the transverse momentum
of the \Dz  as the decaying particle's kinematics are invariant in all redecayed events. On the other hand, quantities computed to evaluate, for example, the
performance of the track reconstruction for the \Km meson tracks will be uncorrelated as those are based on the hits produced by the \Km meson
traversing the detector that are regenerated in every redecayed event.
Additionally, many observables are constrained by the fixed \Dz kinematics but have a certain amount of freedom in each
redecayed event such as the transverse momentum of the \Km. In the following, a simple example is used to develop methods to quantify the degree
of correlation between different events and to discuss methods to take the correlations into account.

Suppose $x$ is a random variable, which is sampled in a two stage process: first, a random number $x'$ is sampled from a normal distribution
$\mathcal{N}(0, \sigma_1)$ with mean zero and width $\sigma_1$. Subsequently, a value for $x$ is then obtained by sampling one number from a normal distribution
$\mathcal{N}(x',\sigma_2)$ with width $\sigma_2$ that is centered at $x'$. The resulting $x$ is then distributed according to the convolution of the 
two normal distributions, $\mathcal{N}(0,\sqrt{\sigma_1^2+\sigma_2^2})$. In the nominal case, a new $x'$ is sampled for every $x$, while a procedure equivalent
to the ReDecay approach is achieved by sampling $N_{\text{ReDecay}}$ values for $x$ from the same $x'$. Modifying the procedure for the generation
of $x$ in this way does not alter the final distribution and thus both approaches provide random numbers that can be used to obtain a histogram following the
expected shape given by the convolution of the two normal distributions.
\begin{figure}
  \begin{subfigure}{0.5\linewidth}
    \includegraphics[width=\linewidth]{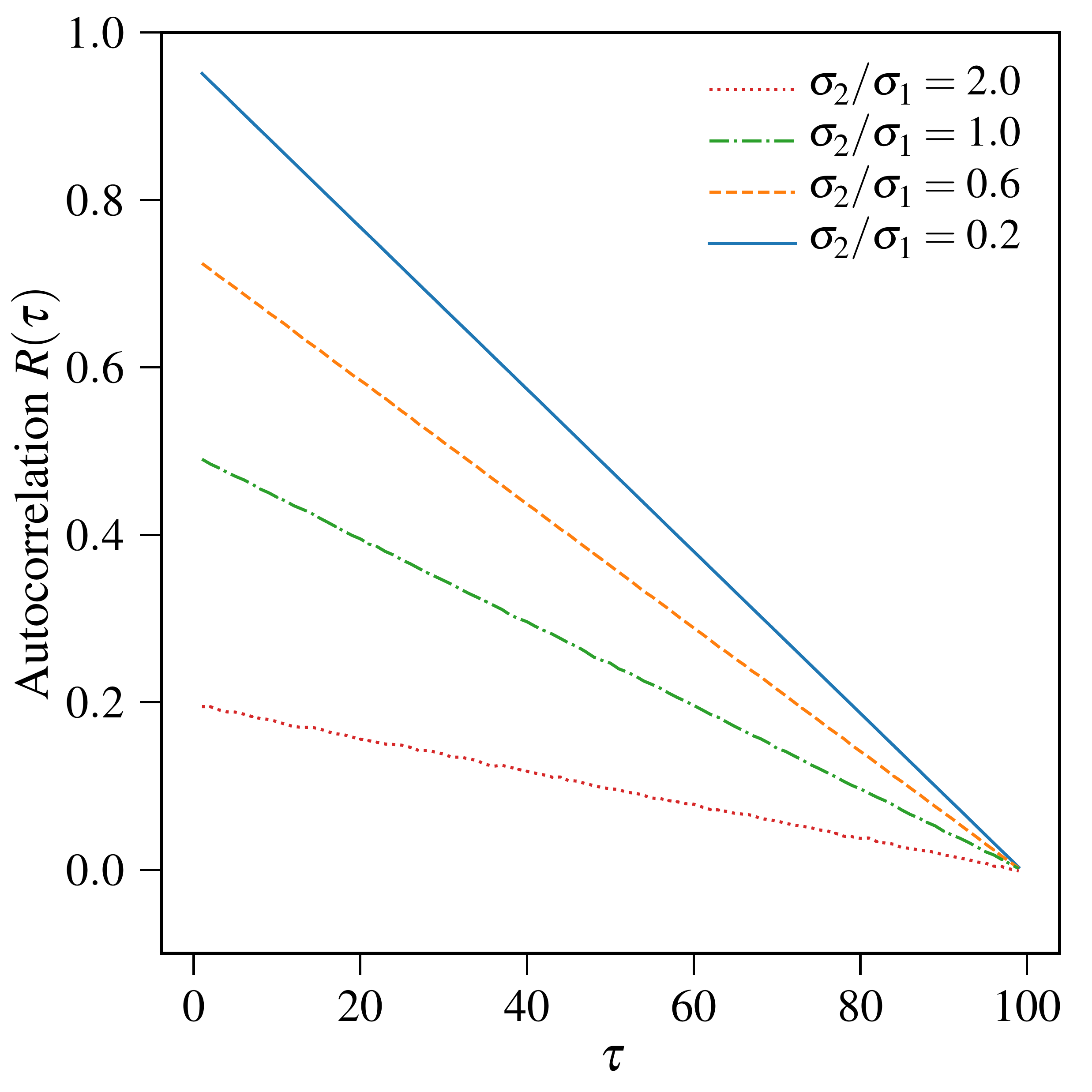}
  \end{subfigure}
  \begin{subfigure}{0.5\linewidth}
    \includegraphics[width=\linewidth, page=1]{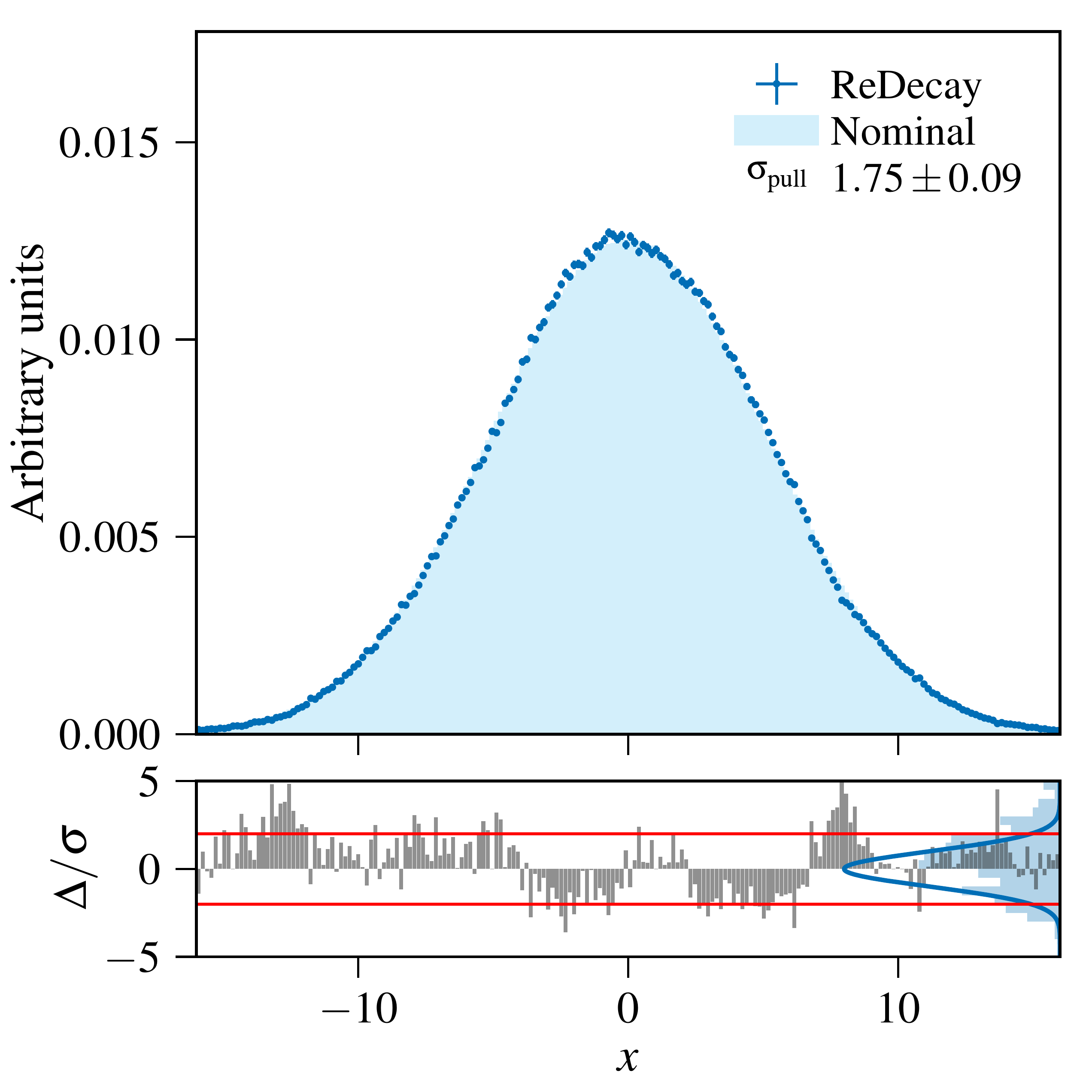}
  \end{subfigure}
  \caption{
    Illustration of the correlations between events produced with the ReDecay method, using $N_{\text{ReDecay}}=100$: (left) autocorrelation function $R(\tau)$ for different values of the ratios of widths of the probability density functions, $\sigma_2/\sigma_1$ and (right) comparison of normalised distributions of a variable $x$ generated with the ReDecay method and a standard generation method (nominal) for $\sigma_1=5$ and $\sigma_2=1$. The bottom part of the right plot shows the pull distribution of the difference $\Delta$ in each bin of the distribution of the value obtained with the ReDecay method with the value from the nominal method, divided by the naive expectation for the statistical undertainty, $\sigma = \sqrt{N}$ where $N$ is the number of events in the bins. A unit Gaussian distribution is overlaid with the pull distribution to visualize their disagreement. The pull distribution has a width of $1.75\pm0.09$ indicating that uncertainties are underestimated in this illustration case.
  }
  \label{fig:illustration}
\end{figure}
However, the latter approach introduces correlations between different $x$ values, which is visible in \cref{fig:illustration} as these correlations lead to an underestimate of the
uncertainties when the values of $x$ are filled into a histogram using uncertainties equal to $\sqrt{N}$ where $N$ is the number of entries in the bin of the nominal distribution. One possibility to quantify the correlation between different entries in a sequence of random numbers is given by the autocorrelation of the random number $x$:

\begin{align}
  R(\tau)=\frac{1}{\left( N-\tau \right)\sigma^2}\displaystyle\sum_{t=1}^{N-\tau}( x_t-\mu ) (x_{t+\tau}-\mu )\, ,
  \label{eq:autocorr}
\end{align}
where $N$ is the total number of entries, $\mu$ is the mean (zero in this example) and $\sigma$ is the standard deviation ($\sqrt{\sigma_1^2+\sigma_2^2}$ here) of the random numbers $x$.
The autocorrelation itself is typically studied as a function of an integer offset $\tau$, where $\tau=0$ is the trivial case with $R(0)=1$. Example autocorrelations as a function
of $\tau$ are given in \cref{fig:illustration} for multiple combinations of values for $\sigma_1$ and $\sigma_2$ with different ratios of $\sigma_2/\sigma_1$. As expected, the autocorrelation decreases both as a function
of the offset $\tau$ as well as with increasing values for the ratio $\sigma_2/\sigma_1$.
While the former is simply caused by a decreasing overlap between the blocks of $x$ from the same $x'$,
the latter is a result of larger values of the ratio $\sigma_2/\sigma_1$ corresponding to a greater amount of freedom in the random process for the generation of $x$ despite using the same $x'$.

In summary, events are correlated though the amount of correlation depends on the studied observable.
This explains the effect seen in the pulls where the approximation of $\sqrt{N}$ for the
uncertainties breaks down in the presence of large correlations, when $\sigma_2/\sigma_1$ is small. In the following, an alternative
approach will be discussed to obtain the statistical uncertainty in the presence of such
correlations.  
The effect of the correlations highly depends on how the sample is used and which variables are
of interest and hence deriving a general, analytical solution is difficult if not impossible.
In use cases where the variables of interest are sufficiently independent in every redecayed event (corresponding to a large $\sigma_2/\sigma_1$ in the example above) even ignoring the correlations altogether can be a valid approach.
Nonetheless, a general solution is provided by bootstrapping \cite{bootstrapping1}, which is a method for generating pseudo-samples by
resampling. These pseudo-samples can then be used to estimate the variance of complex estimators,
for which analytic solutions are impractical.

Starting from a sample with $N$ entries, a pseudo-sample can be obtained according to the following procedure:
first, a new total number of events $N'$ is randomly drawn from a Poisson distribution of mean $N$. Then, entries are
randomly drawn with replacement from the original sample $N'$ times to fill the pseudo-sample.
As sampling with replacement is employed,
 the obtained pseudo-sample can contain some entries of the original sample multiple times while others are not present
 in the pseudo-sample at all.
Unfortunately, this cannot directly be applied to the ReDecay samples because it assumes
statistically independent entries. In the study of time series, data points ordered in time where
each data point can depend on the previous points, different extensions of the bootstrapping
algorithm have been developed to preserve the correlations in the pseudo-samples~\cite{bootstrapping3}.
A common approach is the so-called block bootstrapping where the sample is
divided into blocks. In order to capture the correlations arising in the ReDecay approach,
a block is naturally given by all events using the same ROE (or the same $x'$ in the example above).
To obtain a pseudo-sample, the bootstrapping procedure above is slightly modified:
instead of sampling $N'$ individual entries from the original sample, entire blocks are drawn with replacement
and all entries constituting a drawn block are filled into the pseudo-sample until the pseudo-sample has reached a
size of $N'$ entries.
From these pseudo-samples, derived quantities, e.g. the covariance matrix for the bins in the histogram in \cref{fig:illustration} (right), can be obtained and utilised to include the correlations in an analysis.

\section{Implementation and experience in \lhcb}
\label{sec:experience}
While the idea of ReDecay is applicable in different experiments, the actual implementation strongly depends on the simulation framework and no general solution can be provided.
Nonetheless, this section discusses the experiences gained when introducing the ReDecay approach to the \lhcb software and which can be transferred to other experiments.

In the most commonly used procedure to simulate events in \lhcb, $pp$ collisions are generated with the
\pythia~8.1~\cite{Sjostrand:2006za, Sjostrand:2007gs} generator using a specific \lhcb
configuration similar to the one described in Ref.~\cite{LHCb-PROC-2010-056}.  Decays of particles
are described by \evtgen~\cite{Lange:2001uf} in which final-state
radiation is generated with \photos~\cite{Golonka:2005pn}. The
implementation of the interaction of the generated particles with the detector, and its response,
uses the \geant
toolkit. 
The steering of the different steps in the simulation of an event uses interfaces to external generators and to the \geant toolkit.
It is handled by \gauss~\cite{LHCb-PROC-2011-006}, the \lhcb simulation software built on top of the \gaudi~\cite{gaudi1,gaudi2} framework.

\begin{figure}
  \begin{subfigure}{0.5\linewidth}
    \includegraphics[width=\linewidth, page=1]{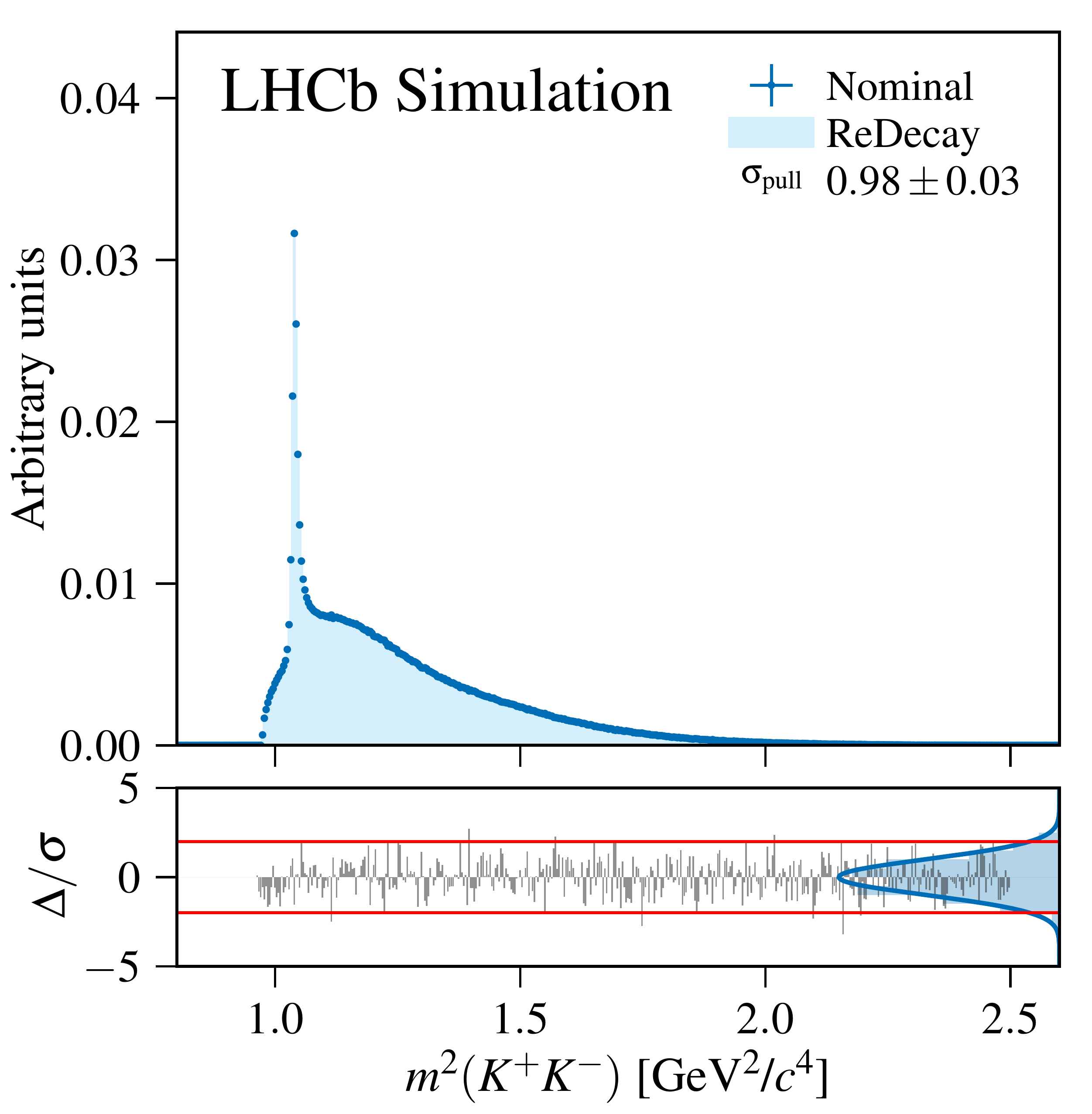}
  \end{subfigure}
  \begin{subfigure}{0.5\linewidth}
    \includegraphics[width=\linewidth, page=2]{./redecay_plots/dtok3pi_plots.pdf}
  \end{subfigure}
  \caption{Comparison of a ReDecay sample with a nominally produced sample of $\Dz\rightarrow\Km\Kp\pip\pim$ decays. Shown are the invariant mass squared of the kaon pair (left) and pion pair (right) which are both independent of the kinematics of the decaying particle. Displayed uncertainties are computed assuming independent events. There is no sign of the effects caused by correlated events as seen in \cref{fig:illustration} (right).}
  \label{fig:dalitzvariables}
\end{figure}

The ReDecay algorithm has been implemented as a package within the \gauss framework, deployed in the
\lhcb production system and already used in several large Monte Carlo sample productions. In general,
a speed-up by a factor of 10 to 20 is observed with the exact factor depending on the simulated conditions
of the Large Hadron Collider and the complexity of the signal event. This factor is reached
with a number of redecays per original event of $N_{\text{ReDecay}}=100$. In this configuration, and depending on the simulated decay, around 95\% of the total time required
to create the sample is spent on the simulation of the detector response of the products of the studied decay.
Hence, increasing the number of redecays further would have little to no impact on the speed.
Due to the correlations between different simulated events using the same ROE as described above,
ReDecay samples are typically used to obtain efficiency descriptions as a function of final state
variables with high granularity.
For example, the use of the ReDecay method is ideal to study variables describing multi-body decays, since these variables are largely independent between all ReDecay events.
An example for this is given in \cref{fig:dalitzvariables}, which compares some distributions
for $\Dz\rightarrow\Km\Kp\pip\pim$ decays between a ReDecay and a nominally simulated sample. Despite ignoring the correlations
and using uncertainties computed as $\sqrt{N}$, sensible pull distributions with a width compatible with one are observed.
In the case where the variables of interest are not obviously independent between different ReDecay events, the degree of correlation
is difficult to predict and the only solution is the actual creation of a test sample. While the increase in speed when using
ReDecay is substantial, the time to produce events is still not negligible and creating a test sample of sufficient size can quickly overwhelm the resources available for an individual analyst. To this end and in collaboration with the respective developers, the ReDecay approach has been added to
the RapidSim~\cite{rapidsim} generator to enable the fast production of simplified simulation samples to judge the degree of correlation before committing
the resources for the production of a large ReDecay sample.

\section{Summary}
The paper presents developments of a fast simulation option applicable to analyses of signal particle decays that occur after the hadronisation phase. An overall increase in speed by a factor of 10 to 20 is observed. Furthermore, this paper discusses procedures to handle the correlations that can arise between different events if necessary. ReDecay has already enabled the production of large simulated samples for current measurements at \lhcb~\cite{LHCb-PAPER-2018-038}.

With the upgrade of the LHCb detector for the third run of the Large Hadron Collider, the detector will collect a substantially larger amount of data. ReDecay is becoming a crucial fast simulation option to extract high precision physics results from this data.

\section*{Acknowledgements}

\noindent We acknowledge all our LHCb collaborators who have contributed to the results presented in this paper. Specifically, we thank Patrick Robbe, Michal Kreps and Mark Whitehead for their editorial help preparing this document. We acknowledge support from the Science and Technology Facilities Council (United Kingdom) and the computing resources that are provided by CERN, IN2P3
(France), KIT and DESY (Germany), INFN (Italy), SURF (The
Netherlands), PIC (Spain), GridPP (United Kingdom), RRCKI and Yandex
LLC (Russia), CSCS (Switzerland), IFIN-HH (Romania), CBPF (Brazil),
PL-GRID (Poland) and OSC (USA). We are indebted to the communities
behind the multiple open-source software packages on which we depend.

\addcontentsline{toc}{section}{References}
\setboolean{inbibliography}{true}
\bibliographystyle{LHCb}
\bibliography{main,LHCb-PAPER,LHCb-CONF,LHCb-DP,LHCb-TDR,additional}

\newpage


 
\newpage

\end{document}